\documentclass[12pt]{iopart}

\usepackage[english]{babel}
\usepackage{revsymb}
\usepackage{amsfonts}
\usepackage{amssymb}
\usepackage{graphicx}

\newcommand{\se}{\textsc{Sec.}~}
\newcommand{\eq}{\textsc{Eq.}~}
\newcommand{\eqs}{\textsc{Eqs.}~}

\newcommand{\fig}{\textsc{Fig.}~}
\newcommand{\figs}{\textsc{Figs.}~}
\begin{document}

\title[]{Dynamics and energetics for a molecular zipper model under
  external driving}

%\title[]{Dynamics and energetics of an extended version of Kittel's
%  molecular zipper model under external driving}

\author{Viktor Holubec$^1$, Petr Chvosta$^1$, Philipp Maass$^{2}$}

\address{$^{1}$ Department of Macromolecular Physics, Faculty of
  Mathematics and Physics, Charles University, CZ-180~00~Praha, Czech
  Republic}

\address{$^{2}$ Fachbereich Physik, Universit\"at Osnabr\"{u}ck,
  Barbarastra{\ss}e 7, 49076 Osnabr\"{u}ck, Germany}
\eads{\mailto{viktor.holubec@gmail.com}}

\date{March 29, 2012}

\begin{abstract}
  We investigate the dynamics of a single-ended $N$-state molecular
  zipper based on a model originally proposed by Kittel. The molecule
  is driven unidirectionally towards the completely unzipped state
  with increasing time $t$. The driving lowers the energies of
  states with $k$ unzipped links by an amount proportional to $kt$. We
  solve the Pauli rate equation for the state probabilities and the
  partial differential equations, which yield the probability
  distributions for the work performed on the zipper and for the heat
  exchanged with the thermal reservoir. Similarly to the related
  equilibrium model, two different regimes can be identified at a
  given temperature with respect to released molecular degrees of
  freedom per broken bond. In these two regimes the time evolution of
  the state probabilities as well as of the work and heat
  distributions show a qualitatively different behavior.
\end{abstract}


\maketitle

\section{Introduction}
\label{sec:introduction}

The Watson-Crick double-stranded form represents the thermodynamically
stable state of DNA in a wide range of temperature and salt
conditions. However, even at standard physiologic conditions, there
always exists possibility that the double-helix is locally unzipped
into two strands both at its ends and in its interior
\cite{Poland_1966, Hanke_2003, Bicout_2004, Fogedby_2007,
  Metzler_2009}.  If the interior unzipping is neglected, the
unfolding of the two strands can be described by a simple zipper model
\cite{Gibbs_1959, Crothers_1965, Kittel_1969}.  This model, while
including several simplifications, emphasizes the essential ingredient
of the unzipping process, that means the competition between the
entropic forces which tend to open the macromolecule, and the
energetic forces that tend to condense it into its double-stranded
form.  

In the last two decades, new experimental techniques have been
developed for detailed analysis of unzipping processes.  A powerful
technique are single molecule manipulations by optical tweezers
\cite{Liphardt_2001, Liphardt_2002, Lang_2003, Onoa_2003, Ritort_2006,
  Ritort_2008}, where biomolecules are unfolded and refolded by
applying mechanical forces. Because single molecules are subject to
strong fluctuations, a stochastic description of the
unfolding/refolding kinetics becomes necessary. Investigation of these
stochastic processes can be useful to understand how biomolecules
unfold and fold under locally applied forces \cite{Tinoco_1999,
  Manosas_2005,Thirumalai_2005, Finkelstein_2004}, as,
for example, when mRNA passes through the ribosome during the
translation process \cite{Green_1997, Ramakrishnan_2002}, or when DNA
is unzipped by helicase during the replication process
\cite{Kornberg_1992, Tackett_2001}.

A particularly interesting part in the analysis of unzipping processes
is the application of integral or detailed fluctuation theorems
\cite{Palassini_2011, Esposito/VandenBroeck:2010,
  VandenBroeck/Esposito:2010, Ritort_2008,Seifert_2008} to estimate
free energy differences between folded and unfolded states (see,
e.g., \cite{Braun_2004}). Their advantage is that they can be applied
also to protocols, which drive the considered system far from
equilibrium.  It is thus not necessary to perform the
unfolding/folding under quasi-static near-equilibrium conditions,
where the process becomes reversible. Most popular theorems are the
Crooks fluctuation theorem \cite{Crooks_1999} and Jarzynski equality
\cite{Jarzynski_1999}. These relate to the distribution of work
performed on the molecule during the process.  Unfortunately it is not
easy to get theoretical insight into characteristics of the underlying
work distributions in far-from-equilibrium processes, since the work
is a functional of the whole stochastic trajectory. Investigations
have been conducted for simple spin systems driven by a time-dependent
external field \cite{Chatelain_2006, Hijar_2007,Chvosta_2007,
  Subrt_2007, Einax_2009,Chvosta_2010} and for diffusion processes in
the presence of a time-dependent potential \cite{Mazonka_1999,
  Baule_2009}.  Analytical solutions are known for two-state systems
\cite{Chvosta_2007, Subrt_2007, Manosas_2009, Chvosta_2010} and for
systems with one continuous state variable \cite{Mazonka_1999,
  Baule_2009}.

In this study we present analytical results for the work distribution
for a multi-state system, which is motivated by a model proposed by
Kittel for describing the melting transition of DNA molecules
\cite{Kittel_1969}. In order to derive analytical results, we had to
consider a stochastic process with directed forwarded and forbidden
backward transitions between states. As a consequence, detailed
balance is broken and fluctuation theorems, as mentioned above, will
not hold true. On the other hand, kinetic Monte Carlo simulations of
the stochastic process show that, for at least certain parameter
settings in the model, the restriction of forbidden backward
transitions is not so severe and events dominating the integrand in
the Jarzynski equality are not very rare. It is important to point out
that these parameter settings are not related to any experimental
conditions. In more realistic settings rare events would play the decisive
role and in such cases it becomes difficult to determine tails of the work
distribution with sufficient accuracy.  Also the relation of the work
considered in our study to the thermodynamic work measured in an
unzipping experiments needs to be treated with care. These problems
are discussed in detail in Sec.~\ref{sec:model} and they imply that
the theory cannot be applied to experiments at this stage. Our
findings should nevertheless be useful because connection to
experiments seems not completely out of reach and because they widen
the range of hopping models, where analytical results for work
distributions are available.

\section{Unzipping in an extended Kittel model}
\label{sec:model}

Kittel's model \cite{Kittel_1969} is a simplified version of the
Poland-Scheraga model \cite{Poland_1966} for the equilibrium
properties of DNA molecules, which got renewed attention in the last
ten years \cite{Lubensky_2000}.  In contrast to the Poland-Scheraga
model, it disregards the possibility of any interior openings
(bubbles) of double-stranded parts of the molecule. Despite this
simplification, it is already sufficient to understand the origin of a
melting transition. 

\begin{figure}[b!]
\center
\includegraphics[width=0.3\textwidth]{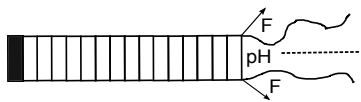}
\caption{The zipper model.}
\label{fig:model}
\end{figure}

A molecule in Kittel's model \cite{Kittel_1969} has $N$ links in its
fully folded state, see \fig\ref{fig:model}. Different states
$k=1,\ldots,N$ of the molecule refer to configurations, where $(k-1)$
of the links are opened in a row. The difference $E_k-E_1$ of the
internal energy of a state with $(k-1)$ open links and the ground
state $k=1$ (no open links) is equal to $(k-1)\Delta$, corresponding
to a loss of chemical bond energy $\Delta$ per broken link. With each
broken link, the molecule gains $G$ degrees of freedom, which
characterize the win of conformational degrees of freedom when
double-stranded DNA is transferred to single stranded DNA. The entropy
difference $S_k-S_1$ between the state $k$ and the ground state then
becomes $k_{\rm B}\ln G^{k-1}= k_{\rm B}(k-1)\ln G$, where $k_{\rm
  B}$ is the Boltzmann constant.  The free energy $F_k$ of state $k$
at a temperature $T$ is thus given by $F_k=F_0+(k-1)[\Delta-k_{\rm
  B}T\ln G]$ and the equilibrium properties can be readily worked out
by considering the partition sum $Z=\sum_{k=1}^N\exp(-\beta F_k)$,
where $\beta=1/k_{\rm B}T$.

In extending this model to treat unfolding kinetics, the molecule is
supposed to unzip successively, one link in each step, as a
consequence of a strong external driving. Such external driving can be
caused by a local force or a pH gradient, as indicated in
\fig\ref{fig:model}.  With respect to pulling experiments on single
DNA molecules out of equilibrium, there is evidence that a neglect of
interior openings can become even less relevant than for the
equilibrium properties. For example, far away from the melting
temperature, e.g.\ at standard room temperature conditions (298~K),
the unfolding kinetics of DNA hairpin molecules could be successfully
described by assuming no interior openings and thermally activated
unzipping transitions \cite{Manosas_2005,Manosas_2009}. Moreover,
molecular fraying can be identified in individual unzipping
trajectories by considering the size of force jumps. Thereby states
with interior openings can be systematically excluded from the
analysis \cite{Engel_2011}.

We assume that the driving lowers the energy differences by an amount
proportional to $(k-1)t$, i.e.\ open states with a larger number of
broken links are favored with increasing time. If we take the ground
state energy as the reference point, $E_1=0$, we obtain for the free
energies $F_k(t)=E_k(t)-k_{\rm B}T\ln G^{k-1}$ of the states
\begin{equation}
\label{eq:Energies}
F_k(t)=(k-1)[\Delta-vt-k_{\rm B}T\ln G]\,, \qquad k = 1,\dots,N\,\,,
\end{equation}
where the parameter $v$ has the dimension of an energy rate
and characterizes different speeds of the unfolding in response to
different strengths of the external driving.

Following established theoretical descriptions for the rupture
kinetics of the bonds \cite{Manosas_2009, Engel_2011, Mossa_2009}, the
Kramers-Bell form \cite{Bell_1978} is used for the time-dependent
transition rate $\lambda_{k,k+1}$ from state $k$ to state $(k+1)$,
\begin{equation}
\lambda_{k,k+1}(t)=\tilde{\nu}\exp[-\beta F_{k,k+1}(t)]\,.
\label{eq:lambda}
\end{equation}
Here $\tilde{\nu}$ is an attempt frequency and $F_{k,k+1}(t)$ denotes
the free energy barrier for the transition, i.e.\ the difference of
the free energy at the saddle point separating states $k$, $(k+1)$ and
the free energy $F_k$ in state $k$.  The barrier $F_{k,k+1}(t)$ is
considered to be composed of a bare, $k$-independent free energy
barrier, $F_b$, which is modified by an amount proportional to
the free energy difference, $[F_{k+1}(t)-F_k(t)]$,
\begin{equation}
\hspace*{-1cm}F_{k,k+1}(t)=F_b+\gamma [F_{k+1}(t)-F_k(t)]=F_b+
\gamma[\Delta-vt-k_{\rm B}T\ln G]\,\,.
\label{eq:deltaf}
\end{equation}
In the following we set $\gamma=1$ \cite{comm:gamma-dfb}. Note that
$F_{k,k+1}(t)$ and thus $\lambda(t)\equiv \lambda_{k,k+1}(t)$ in
\eq(\ref{eq:lambda}) are independent of $k$. The
attempt frequency $\tilde\nu$ in \eq(\ref{eq:lambda}) was reported
\cite{Cocco_2001} to be approximately proportional to the ratio
of the diffusion constant of the molecule in water to the water
viscosity. We assume here a linear dependence of this ratio on the
temperature $T$, i.e.\ we take $\tilde{\nu} = \nu(T/T_0)$, where $\nu$
and $T_0$ are positive constants. Thus we obtain
\begin{equation}
\label{eq:rates}
\lambda(t)=\lambda_{k,k+1}(t)=g\exp[-\beta(d-vt)]\,\,,
\end{equation}
where
\begin{equation}
\label{eq:g-b}
g=\nu G\frac{T}{T_0}\,,\qquad d=\Delta+F_b\,.
\end{equation}

If detailed balance is obeyed, i.e.\ $\lambda_{k,k+1}(t) \exp[- \beta
F_k(t)] = \lambda_{k+1,k}(t) \exp[- \beta F_{k+1}(t)]$, the rates
$\lambda_{k+1,k}(t)\equiv\lambda_b$ for backward (refolding)
transitions become both independent of $k$ and $t$,
\begin{equation}
\lambda_b=\lambda_{k+1,k}(t) = \frac{g}{G}\exp(-\beta F_b)\,\,.
\label{eq:back_rates}
\end{equation}
Note that $G$ appears here in the denominator, which means that the
ratio $\lambda_b/\lambda(t)=\exp(\beta\Delta-\beta vt)/G$ of backward to
forward rates becomes small for large degeneracy factors $G$. This
reflects the fact that it is difficult for the flexible unfolded part
of the molecule to find proper configurations, which would allow for a
reformation of (hydrogen) bonds.

If the model would refer to the hopping motion of a particle between
time-dependent energy levels $E_k(t)$, the work performed on the
system (for one realization of the stochastic process) would be
\begin{eqnarray}
\mathsf{W}(t)&=&\int_0^t{\rm d}t'\,\sum_{k=1}^N \dot E_k(t')
\delta_{k \mathsf{D}(t')}\nonumber\\
&=&[E_{\mathsf{D(t)}}(t)-E_{\mathsf{D(t)}}(t_{\mathsf{D(t)}})]+
\sum_{k=1}^{\mathsf{D}(t)-1}[E_k(t_{k+1})-E_k(t_k)]\nonumber\\
&=&[F_{\mathsf{D(t)}}(t)-F_{\mathsf{D(t)}}(t_{\mathsf{D(t)}})]+
\sum_{k=1}^{\mathsf{D}(t)-1}[F_k(t_{k+1})-F_k(t_k)]\,\,,
\label{eq:work}
\end{eqnarray}
where $\mathsf{D}(t)$ denotes the (random) state of the molecule at
time $t$, $\dot E_k(t)={\rm d}E_k(t)/{\rm d}t$, and $t_k$ is the
(random) time at which the transition from state $k$ to $(k+1)$
(rupture of $k$th bond) takes place ($t_1=0$ being the initial time).
The last line in \eq(\ref{eq:work}) follows from the fact that the
entropy in the extended Kittel model is not dependent on time, i.e.,
for given $k$, differences between internal energies and between free
energies at distinct times are equal.

The work in \eq(\ref{eq:work}) can be related to the thermodynamic
work $W_f$ in an unzipping experiment under force control. As pointed
out in \cite{Alemany_2011}, work in thermodynamics is the internal
energy transferred to a system upon changing the control parameters
for given system configuration. This means that, when the force
$f=f(t)$ is the control variable and the molecular extension $x$ the
conjugate configurational variable, one has
$W_f(t)=-\int_{f(0)}^{f(t)}{\rm d}fx$ \cite{comm:tweezers}.
Stochastic changes of the molecular extension occur mainly due to bond
rupture, while in between transitions the response will be rather
smooth and, under neglect of small thermal fluctuations, can be
represented by a function $x(k,f)$. This specifies the {\it mean}
end-to-end distance of the unfolded part in state $k$ at force $f$ (as
often modeled by, e.g., the freely jointed or worm-like chain models
from polymer physics). With $f_k\equiv f(t_k)$ we then have
\begin{equation}
\label{eq:work-f}
\hspace*{-1cm}W_f(t)=-\int_{f(0)}^{f(t)}{\rm d}f\,x=
-\int_{f_{\mathsf{D}(t)}}^{f(t)}{\rm d}f\, x(\mathsf{D}(t),f)
-\sum_{k=1}^{\mathsf{D}(t)-1}
\int_{f_k}^{f_{k+1}}{\rm d}f\, x(k,f)\,\,.
\end{equation}

In more detailed energy landscape models, the free energy $F_{\rm
  tot}(k,f)$ of the molecule in state $k$ under loading $f$ can be
represented as $F_{\rm tot}(k,f)=F_0(k)-fx(k,f)+F_{\rm str}(k,f)$
(see, e.g., \cite{Manosas_2005}), where $F_0(k)$ is the free energy in
the absence of loading and $F_{\rm str}(k,f)$ the elastic energy of
the unfolded part upon stretching. For smooth response under
stretching the latter is given by \cite{Manosas_2005}
\begin{equation}
\label{eq:e-str}
F_{\rm str}(k,f)=\int_{x(k,0)}^{x(k,f)}{\rm d}x\, \tilde f(k,x)=
fx(k,f)-\int_0^fdf'\,x(k,f')\,,
\end{equation}
where $\tilde f(k,x)$ denotes the inverse function of $x(k,f)$ with
respect to $f$. Accordingly, at a given $k$, the work for stretching 
upon increasing the force from $f_a$ to $f_b$ becomes
\begin{eqnarray}
-\int_{f_a}^{f_b}df\,x(k,f)&=&
F_{\rm str}(k,f_b)-f_bx(k,f_b)-F_{\rm
  str}(k,f_a)+f_ax(k,f_a)\nonumber\\
&=&F_{\rm tot}(k,f_b)-F_{\rm tot}(k,f_a)\,\,.
\label{eq:wstr-ftot}
\end{eqnarray}
This just means that the stretching at fixed $k$ is assumed to
take place quasi-statically, i.e.\ the variation of the force is
supposed to occur on a time scale much slower than the
correlation time of end-to-end distance fluctuations. Inserting this
result into \eq(\ref{eq:work-f}) yields
\begin{equation}
\label{eq:work-f-2}
\hspace*{-2cm}W_f(t)=-[F_{\rm tot}(k,f(t))-F_{\rm tot}(k,\mathsf{D}(t))]
-\sum_{k=1}^{\mathsf{D}(t)-1}[F_{\rm tot}(k,f(t_{k+1}))-F_{\rm tot}(k,f(t_k))]\,,
\end{equation}
from which it becomes clear that $W_f$ equals $(-W)$, if $F_{\rm
  tot}(k,f(t))$ is identified with the free energy $F_k(t)$ in the
extended Kittel model.

As said in the Introduction, we were able to find analytical results
for the work distributions, if the backward rates in
\eq(\ref{eq:back_rates}) were negligible. We are interested here in
the model itself and will not make attempts to assign values to the
parameters in the transition rates $\lambda(t)$ [\eq(\ref{eq:rates})]
and $\lambda_b$ [\eq(\ref{eq:back_rates})], and appearing also in the
state free energies in \eq(\ref{eq:Energies}), which are connected
to real experiments. Nevertheless, with respect to the value of the
analytical results in connection with fluctuation theorems, the
question arises whether a neglect of backward rates could be
acceptable, at least for certain parameter settings. To check this, we
have performed kinetic Monte Carlo simulations of the stochastic
process.  Because the forward rate from \eq(\ref{eq:rates}) can be
written as $\lambda(t)=\lambda_bG\exp(-\beta\Delta+\beta vt)$ and the
free energy differences appearing in \eq(\ref{eq:work}) by
$[F_k(t_{k+1})-F_k(t_k)]=-(k-1)v(t_{k+1}-t_k)$, the dynamics and
energetics of the model is completely specified by the parameters
$\lambda_b$, $G$, $v$, and $T$ (and $N$ if we consider a complete
unfolding). For illustration and discussion of representative results,
we here use $\Delta$, $\Delta/k_{\rm B}$, and $\nu^{-1}$ as units for
energy, temperature and time, respectively.

\begin{figure}
\centering
\includegraphics[width=0.9\textwidth]{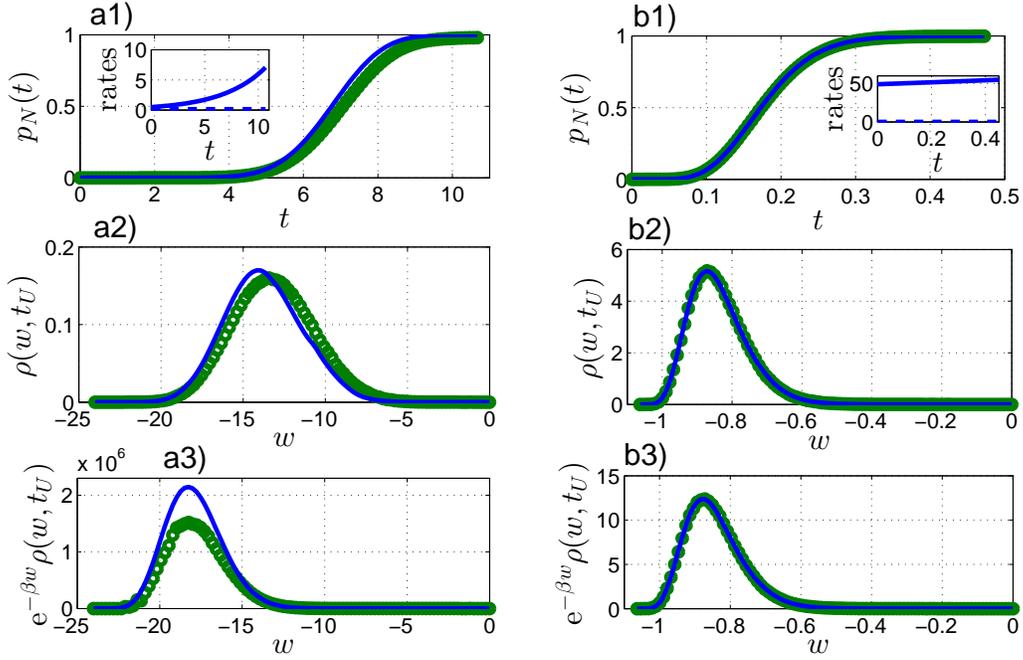}
\caption{Simulated results of the model presented in
  \se\ref{sec:model} with backward transition rate $\lambda_b=0.133$
  (blue lines) in comparison with analytical results (see
  Sec.~\ref{sec:solution}) when neglecting backward transitions (green
  circles). Panels labeled with a) and b) refer to degeneracy
  factors $G=10$ and $G=1000$, respectively. The remaining parameters
  are $T=1$, $v=0.25$, and $N=10$. Panels a1) and b1) show the
  probability $p_N(t)$ that the zipper has unfolded up to time $t$ and
  the insets depict the time-dependence of the forward transition
  rates (full line); the constant backward rate is indicated by the
  dashed line. Panels a2) and b2) display the probability densities
  $\rho(w,t_U)$ at the unfolding time $t_U$ ($t_U=10.67$ for $G=10$
  and $t_U=0.47$ for $G=1000$), and panels a3) and b3) the weighted
  probability densities $\exp(-\beta w)\rho(w,t_U)$.}
\label{fig:backward_rates}
\end{figure}

Simulations were performed for fixed $N=10$, $T=1$ and $\lambda_b=1$,
and a set of $v$ and $G$ values varying in the intervals $v=0.01-3.3$
and $G=10-1000$, respectively. We always started the unfolding from
the fully closed state, i.e.\ $p_k(0)=\delta_{k,1}$
\cite{comm:modif-jarz}.  Probabilities $p_N(t)$ of complete unfolding
(occupation of state $N$) until time $t$ were determined and an
unfolding time $t_U$ defined by requiring that at $t=t_U$ the zipper
has unfolded with a probability of 99.9\%, i.e.\ $p_N(t_U)=0.999$ (see
Sec.~\ref{sec:discussion}). We then considered the work distributions
$\rho(w,t_U)$ at $t=t_U$ and the weighted distributions $\exp(-\beta
w)\rho(w,t_U)$, corresponding to the integrand in the average
$\langle\exp(-\beta w)\rangle=\int {\rm d}w\,\exp(-\beta
w)\rho(w,t_U)$ as it appears in the Jarzynski equality. It was found
that backward rates turn out to have a minor importance when the
forward rates are much larger than the backward rates during the whole
unfolding process or when $v$ is large enough. Specifically, the error
in calculating $\langle \exp(-\beta w)\rangle$ is smaller than
5\% when $v\gtrsim 3$ for $G=10$, $v\gtrsim 0.1$ for $G=100$, and
$v\gtrsim0.01$ for $G=1000$. As representative examples we show in
\fig\ref{fig:backward_rates} simulated results (blue lines) with
nonzero backward rates in comparison with analytical results (green
circles) for zero backward rates (see Sec.~\ref{sec:solution}) for
$p_N(t)$, $\rho(w,t_U)$, and $\exp(-\beta w)\rho(w,t_U)$, and
parameters $v=0.25$, $G=10$ [panels labeled with a)] and $G=1000$
[panels labeled with b)].  As can be seen from the figure, for the case
of large $G$ (small backward transitions), the simulated results for
$\exp(-\beta w)\rho(w,t_U)$ are almost indistinguishable from the
analytical results.

% \fig\ref{fig:backward_rates} shows simulated results (blue lines)
% in comparison with analytical results (green circles) for zero backward
% rates (see Sec.~\ref{sec:solution}) for probabilities $p_N(t)$ of
% complete unfolding (occupation of state $N$) until time $t$ and for
% work distributions $\rho(w,t_U)$. The time $t_U$ refers to ``complete
% opening'', which we define by saying that the zipper has unfolded with
% a probability of 99.9\%, i.e.\ $p_N(t_U)=0.999$ (see
% Sec.~\ref{sec:discussion}). Also shown are results for $\exp(-\beta
% w)\rho(w,t_U)$, corresponding to the integrand in the average
% $\langle\exp(-\beta w)\rangle=\int {\rm d}w\,\exp(-\beta
% w)\rho(w,t_U)$ as it appears in the Jarzynski equality. Two parameter
% sets are considered, which differ only in the degeneracy
% factor. Panels labeled with a) are for $G=10$, while panels labeled
% with b) are for $G=1000$.  We always started the unfolding from the
% fully closed state, i.e.\ $p_k(0)=\delta_{k,1}$
% \cite{comm:modif-jarz}.  As can be seen in the figure, for the case of
% large $G$ (small backward transitions), the simulated results for
% $\exp(-\beta w)\rho(w,t_U)$ agree with the analytical results.

The remaining part of the paper is organized as follows. 
In \se\ref{sec:evolution} we specify the equations for the time
evolution of the state probabilities [\eq(\ref{eq:Pauli})] and for the
quantities describing the energy transformations
[\eq(\ref{eq:G_equation})]. In \se\ref{sec:solution} we derive exact
analytical solutions of these equations and in \se\ref{sec:discussion}
we discuss our findings.

%********************************************************************
%*********************************************************************

\section{Time evolution of state probabilities and work}
\label{sec:evolution}

Let $p_k(t)$, $k=1,\dots,N$, be the occupation probabilities of the
$k$-th state. The time evolution of these functions is governed by the
Pauli rate equation with transition rates given by
\eq(\ref{eq:rates}).  Formally speaking, the unzipping process is
described by the time-inhomogeneous Markov process $\mathsf{D}(t)$,
where $\mathsf{D}(t)=k$ if the system resides in state $k$ at time
$t$. The Pauli rate equation can be written as
\begin{equation}
	\label{eq:Pauli}
	\frac{\mathrm{d} }{\mathrm{d} t}\mathbb{R}(t)=
\mathbb{L}(t)\,\mathbb{R}(t)\,\,,\qquad \mathbb{R}(0) = \mathbb{I} \,\,,
\end{equation}
where $\mathbb{I}$ is the $(N\times N)$ unity matrix, $\mathbb{L}(t)$
is the $(N\times N)$ matrix of the transition rates,
\begin{equation}
\label{eq:Rate_matrix}
\mathbb{L}(t)=\left(
\begin{array}{ccccc}
-\lambda(t) & 0 &  \dots & \dots  & 0\\
 \lambda(t) & - \lambda(t) & \ddots  &   & \vdots  \\
 0 & \ddots & \ddots & \ddots  & \vdots \\
 \vdots &  \ddots  & \ddots & - \lambda(t) & 0  \\
 0 & \dots  &  0  & \lambda(t) & 0
\end{array}
\right) \,\,,
\end{equation}
and ${\mathbb R}(t)$ is the $(N\times N)$ matrix of the transition
probabilities with the matrix elements
\begin{equation}
\label{eq:Pauli_elements}
R_{ij}(t)={\rm Prob}\left\{\,\mathsf{D}(t)=i\,|\,\mathsf{D}(0)=j\,\right\}\,\,.
\end{equation}
The matrix of transition probabilities evolves an arbitrary column
vector of the initial occupation probabilities,
$p(t)=\mathbb{R}(t)\,p(0)$. In the following, we always start with the
completely closed zipper, that means $p_{k}(0)=\delta_{k 1}$.  The
individual occupation probabilities then are
\begin{equation}
\label{eq:Probabilities}
p_i(t) = R_{i1}(t)\,\,.
\end{equation}

According to \eq(\ref{eq:work}) the work $\mathsf{W}(t)$ is a {\em
  functional\/} of the process $\mathsf{D}(t)$. For an analytical
treatment it is useful to introduce the \emph{augmented process}
$\left\{\mathsf{W}(t),\,\mathsf{D}(t)\right\}$ \cite{Imparato_2005A,
  Imparato_2005B, Subrt_2007} which describes both the work and the
state variable. This augmented process is again a time non-homogeneous
Markov process and its one-time properties are described by $(N\times
N)$ matrix ${\mathbb G}(w,t)$ with the matrix elements
\begin{equation}
\fl \label{eq:G_elements}
G_{ij}(w,t)=\lim \limits_{\epsilon \to 0}
\frac{ {\rm Prob}\left\{\,\mathsf{W}(t)\in(w,w+\epsilon)\,{\rm and}\,
\mathsf{D}(t)=i\,|\,\mathsf{W}(0)=0\,{\rm and}\,\mathsf{D}(0)=j\,\right\}}
{\epsilon}\,\,.
\end{equation}
The time evolution of ${\mathbb G}(w,t)$ is given by
\cite{Imparato_2005A, Imparato_2005B, Subrt_2007}
\begin{equation}
\label{eq:G_equation}
\frac{\partial}{\partial t}{\mathbb
G}(w,t)= \left[- \frac{\partial}{\partial w}\, \,\dot{{\mathbb E}}(t)
+\mathbb{L}(t)\,\right]\,{\mathbb G}(w,t)\,\,,
\qquad {\mathbb G}(w,0)=\delta(w)\,{\mathbb I}\,\,.
\end{equation}
Here $\dot{{\mathbb E}}(t)$ is the diagonal matrix $\dot{{\mathbb
    E}}(t) ={\rm diag}\{\dot{E}_1(t),\dots,\dot{E}_{N}(t)\}$. Notice
that ${\mathbb R}(t)=\int_{-\infty}^{\infty}{\rm d}w\,{\mathbb
  G}(w,t)$. \eq(\ref{eq:G_equation}) represents a hyperbolic system of
$N^{2}$ coupled partial differential equations with time-dependent
coefficients.  Its exact solution will be given in the following
\se\ref{sec:solution}.

The matrix ${\mathbb G}(w,t)$ provides a complete description of the
energetics of the unzipping process. The joint probability density for
the internal energy $\mathsf{U}(t)=E_{\mathsf{D}(t)}(t)$ and the work
$\mathsf{W}(t)$ performed on the system during the time interval
$[0,t]$ (regardless of the final state of the system at the time $t$)
is given by
\begin{equation}
\label{eq:UWPD}
\xi(u,w,t)=\sum_{i=1}^N\, \delta\left[u - E_i(t)\right]\, G_{i1}(w,t)\,\,,
\end{equation}
where $\delta(x)$ is the Dirac $\delta$-function. The last function
already yields the probability density for the work,
\begin{equation}
\label{eq:WPD}
\rho(w,t) = \int_{-\infty}^{\infty}{\rm d}u\,\,\xi(u,w,t)\,\,.
\end{equation}
An analogous integration over the work variable $w$ gives the
probability density of $\mathsf{U}(t)$. Furthermore, the first law of
thermodynamics implies
$\mathsf{U}(t)-\mathsf{U}(0)=\mathsf{U}(t)=\mathsf{W}(t)+\mathsf{Q}(t)$
(note that $\mathsf{U}(0)=0$ for our setting) and, accordingly,
$\xi(u,w,t)$ gives also the probability density of the heat
$\mathsf{Q}(t)$ transferred from the reservoir during the time
interval $[0,t]$ \cite{Chvosta_2010}.

The mean values of the internal energy, work and heat, are then
\begin{eqnarray}
\label{eq:Umean}
\rule[-1ex]{0em}{6ex}U(t)&=&
\int_{-\infty}^{\infty}\int_{-\infty}^{\infty}{\rm d}u\,{\rm d}w\,u\,\xi(u,w,t)
= \sum_{i=1}^N E_i(t) p_i(t)\,\,,\\
\label{eq:Wmean}
\rule[-1ex]{0em}{6ex}W(t)&=&
\int_{-\infty}^{\infty}\int_{-\infty}^{\infty}{\rm d}u\,{\rm d}w\,w\,\xi(u,w,t)
= \sum_{i=1}^N \int_0^t{\rm d}t\,\dot{E}_i(t) p_i(t)
\,\,,\\
\label{eq:Qmean}
\rule[-1ex]{0em}{6ex}Q(t)&=&
\int_{-\infty}^{\infty}\int_{-\infty}^{\infty}{\rm d}u\,{\rm d}w\,(u - w)\,\xi(u,w,t)
= \sum_{i=1}^N \int_0^t{\rm d}t\,E_i(t) \dot{p}_i(t)  \,\,.
\end{eqnarray}
Notice that these mean values can be also calculated directly from the
solution of the Pauli rate equation (\ref{eq:Pauli}). However, for
higher moments, we already need the function $\xi(u,w,t)$.  For
example, the variances discussed in \se\ref{sec:discussion} are given
by
\begin{eqnarray}
\label{eq:Uvar}
\rule[-1ex]{0em}{6ex}[\Delta U(t)]^2&=&
\int_{-\infty}^{\infty}\int_{-\infty}^{\infty}{\rm d}u\,{\rm d}w\,u^2\,\xi(u,w,t)
- [U(t)]^2  \,\,,\\
\label{eq:Wvar}
\rule[-1ex]{0em}{6ex}[\Delta W(t)]^2&=&
\int_{-\infty}^{\infty}\int_{-\infty}^{\infty}{\rm d}u\,{\rm d}w\,w^2\,\xi(u,w,t)
- [W(t)]^2   \,\,,\\
\label{eq:Qvar}
\rule[-1ex]{0em}{6ex}[\Delta Q(t)]^2&=&
\int_{-\infty}^{\infty}\int_{-\infty}^{\infty}{\rm d}u\,{\rm d}w\,(u - w)^2\,
\xi(u,w,t) - [Q(t)]^2  \,\,.
\end{eqnarray}

\section{Solution of the model}
\label{sec:solution}
Owing to the simple two-diagonal structure of the matrix
$\mathbb{L}(t)$ in \eq(\ref{eq:Rate_matrix}), the system
(\ref{eq:Pauli}) can be solved by simple integrations. By defining
\begin{equation}
\label{eq:Int_rate}
\Lambda(t,t')=\int_{t'}^t\,{\rm d}t''\lambda(t'')
=\frac{\alpha}{\beta v} \left[\exp(\beta v t)-\exp(\beta v t')\right]\,\,,
\end{equation}
where $\alpha=g\exp(-\beta d)$, a recursive treatment of
\eq(\ref{eq:Pauli}) yields
\begin{eqnarray}
\label{eq:Sol_Pauli}
\begin{array}{lc}
\rule[-1ex]{0em}{6ex}R_{i j}(t) = 0\,\,,&\quad i < j \,\,,\\
\rule[-1ex]{0em}{6ex}R_{j j}(t) =
\exp{\left[ - \Lambda(t,0) \right]}\,\,, &\quad j=1,\ldots,N\,\,, \\
\rule[-1ex]{0em}{6ex}R_{i j}(t) =
\displaystyle{\int_{0}^t}\,{\rm d}t'
\exp{\left[ - \Lambda(t,t') \right]}\,\lambda(t')\,R_{i - 1 j}(t') \,\,,
&\quad j < i < N\,\,,\\
\rule[-1ex]{0em}{6ex}R_{N j}(t) =
\displaystyle{\int_{0}^t}\,{\rm d}t' \lambda(t')\, R_{N - 1 j}(t')\,\,,
&\quad j=1,\ldots,N\,\,.
\end{array}
\end{eqnarray}
When solving these recursive relations, we obtain the lower triangular
matrix with the nonzero matrix elements
\begin{eqnarray}
\label{eq:R_small}
\fl\rule[-1ex]{0em}{6ex}R_{j + k j}(t)&=&
\frac{\left[\Lambda(t,0)\right]^k}{k!}\,
\exp\left[-\Lambda(t,0)\right]\,\,,\,\,\quad
j=1,\ldots,N-1\,\,;\,\,k=0,\dots,N - 1 -j\,\,,\\
\label{eq:R_big}
\fl\rule[-1ex]{0em}{6ex}R_{N j}(t)&=&
\frac{\left[\Lambda(t,0) \right]^{N - j}}{(N - j)!}\,
\exp\left[ - \Lambda(t,0) \right]\,{}_1\!F_1(1, N + 1 - j;\Lambda(t,0))
\,\,,\,\,\quad j=1,\ldots,N\,\,.
\end{eqnarray}
where ${}_1\!F_1(a, b;x)$ denotes the confluent hypergeometric
function.

For solving \eq(\ref{eq:G_equation}) we first perform a Laplace
transform with respect to the work variable $w$. To keep the notation
simple, we use the same symbols for the original functions and for the
transformed ones. The transformed functions will be distinguished by
explicitly giving the complex variable $s$ conjugate to $w$.  After
performing the Laplace transformation, we get the system of ordinary
differential equations
\begin{equation}
\label{eq:Green_WPD}
\frac{\partial}{\partial t}{\mathbb G}(s,t)=
\left[-s\,\dot{{\mathbb E}}(t)+\mathbb{L}(t)\,\right]\,{\mathbb G}(s,t)\,\,,
\qquad {\mathbb G}(s,0)={\mathbb I}\,\,.
\end{equation}
The matrix which multiplies ${\mathbb G}(s,t)$ on the
right hand side is again a lower two-diagonal one. Therefore,
similarly to \eq(\ref{eq:Sol_Pauli}), we find the recursive relation
\begin{equation}
\fl\label{eq:Sol_Green_WPD_s}
\begin{array}{lc}
\rule[-1ex]{0em}{6ex}G_{i j}(s,t) = 0\,\,,&\quad i < j \,\,,\\
\rule[-1ex]{0em}{6ex}G_{j j}(s,t) =
\exp{\left[ - \Lambda(t,0) \right]}
\exp{\left \{- s [E_j(t) - E_j(0)]\right\}}\,\,,&\quad j=1,\ldots,N \,\,, \\
\rule[-1ex]{0em}{6ex}G_{i j}(s,t) =
\displaystyle{\int_0^t}\,{\rm d}t' \exp{\left[ - \Lambda(t,t') \right]}
\exp{\left \{ - s [E_i(t) - E_i(t')]\right\}}\,\lambda(t')\,
G_{i - 1 j}(s,t')\,\,,&\quad j < i < N \,\,,\\
\rule[-1ex]{0em}{6ex}G_{N j}(s,t) =
\displaystyle{\int_0^t}\,{\rm d}t'
\exp{\left \{- s [E_{N}(t) - E_{N}(t')]\right\}}\,\lambda(t')\,
G_{N - 1 j}(s,t')\,\,,&\quad j=1,\ldots,N \,\,.
\end{array}
\end{equation}
Notice that the matrix ${\mathbb G}(s,t)$ is again a lower triangular
one. We now want to solve these recursive relations. It turns out that
all matrix elements of the matrix ${\mathbb G}(s,t)$, except the
matrix elements $G_{N j}(w,s)$, $j=1,\ldots,N-1$, can be explicitly
evaluated by simple integrations:
\begin{equation}
\fl \label{eq:Sol_Green_WPD_s1}
G_{j + k j}(s,t)=\frac{\left\{\displaystyle\frac{\alpha}{v(\beta - s)}
\left[\exp(\beta v t) - \exp(s v t)\right]\right\}^k}{k!}\,
\exp\left[s(j -1)vt\right]\,\exp{\left[-\Lambda(t,0) \right]}\,\,,
\end{equation}
for $j=N$, $k=0$ and also for $j=1,\ldots,N-1$, $k=0,\dots,N - 1
-j$. Moreover, we were able to carry out the inverse Laplace
transformation of these functions. The resulting diagonal elements
$G_{j j}(w,t)$, $j=1,\ldots,N$ are proportional to Dirac
$\delta$-functions. The remaining ones possess a finite support, i.e.\
they are proportional to the differences of the unit-step functions
$\Theta(a,b;x)=\Theta(x-a)-\Theta(x-b)$,
\begin{eqnarray}
\label{eq:Sol_Green_WPD_w1}
\fl\rule[-1ex]{0em}{6ex}{\phantom +}G_{j j}(w,t)&=&
\delta[w + (j-1)vt]\,\exp{\left[-\Lambda(t,0) \right]}
\,\,,\,\,j=1,\ldots,N\,\,,\\
\fl\rule[-1ex]{0em}{6ex}G_{j + k j}(w,t)&=&
\displaystyle\exp{\left[-\Lambda(t,0)\right]}\,
\frac{\displaystyle\left(-\frac{\alpha}{v}\right)^k}{(k-1)!}\,
\exp\left\{\beta[w + (k + j - 1)vt]\right\}\times\nonumber\\
\fl\label{eq:Sol_Green_WPD_w2}
\rule[-1ex]{0em}{6ex}&\times&\sum_{l=0}^k
\frac{(-1)^{l}}{l!\,(k-l)!}\,
\frac{\Theta[(1 - j - l)vt,(1 -j)vt;w]}{\left[ w + (j+l-1)vt \right]^{1 - k}}
\,\,.
\end{eqnarray}
These expressions are valid for $j=1,\ldots,N-1$ and $k=1,\dots,N - 1
-j$.

It remains to calculate the matrix elements $G_{N j}(s,t)$,
$j=1,\ldots,N$. The inverse Laplace transformation of the last
equation in the recursive scheme (\ref{eq:Sol_Green_WPD_s}) is
\begin{equation}
\fl\label{eq:Sol_Green_WPD_w3}
G_{N j}(w,t)=\displaystyle{\int_0^t}\,{\rm d}t' \lambda(t')\,
G_{N - 1 j}\left\{w - [E_{N}(t) - E_{N}(t')],t'\right\}
\,\,,\,\,j=1,\ldots,N-1\,\,.
\end{equation}
We insert herein the explicit forms of \eqs(\ref{eq:Sol_Green_WPD_w1})
and (\ref{eq:Sol_Green_WPD_w2}). After some algebra we finally obtain
\begin{eqnarray}
\label{eq:Sol_Green_WPD_w4}
\fl\rule[-1ex]{0em}{6ex}G_{N N - 1}(w,t)&=&
\frac{\alpha}{v}\,\exp\left\{\beta[w+(N-1)vt]\right\}\,
\exp{\left\{-\Lambda\left[\displaystyle\frac{w}{v}+(N-1)t,0\right]\right\}}
\times\nonumber\\
\fl\rule[-1ex]{0em}{6ex}&\times&\Theta[-(N-1)vt,-(N-2)vt;w]\,\,,\\
\label{eq:Sol_Green_WPD_w5}
\fl\rule[-1ex]{0em}{6ex}G_{N j}(w,t)&=&
\frac{\displaystyle\alpha\left(-\frac{\alpha}{v}\right)^{N-1-j}}{(N-2-j)!}\,
\exp\{\beta[w+(N-1)vt]\}
\,\sum_{l=0}^{N-1-j}\frac{(- 1)^{l}}{l!\,(N-1-j-l)!}\,
\times\nonumber\\
\fl\rule[-1ex]{0em}{6ex}&\times&
\Bigg\{F_{jl}\left[\frac{w + (N-1)vt}{v(N-j)},
\frac{w + (N-1)vt}{v(N-j-l)};w,t\right]\,
\Theta[-(N-1)vt,-(j + l - 1)vt;w]+\nonumber\\
\fl\rule[-1ex]{0em}{6ex}&+&
F_{jl}\left[\frac{w + (N-1)vt}{v(N-j)},t;w,t\right]\,
\Theta[-(j + l-1)vt,-(j-1)vt;w]\Bigg\}\,\,,
\end{eqnarray}
Here $j \in \{1,\dots, N-2\}$ and we have introduced the abbreviation
\begin{equation}
\label{eq:Sol_Green_WPD_w6}
\fl F_{jl}(a,b;w,t)=\int_{a}^{b}{\rm d}t'\,
\exp{\left[-\Lambda\left(t',0\right)\right]}\,
\left[ w + (N-1)vt - (N-j-l)vt' \right]^{N-j-2} \,\,.
\end{equation}
The main results of this Section are \eqs(\ref{eq:R_small}) and
(\ref{eq:R_big}), which give the solution of the Pauli rate equation
(\ref{eq:Pauli}), and
\eqs(\ref{eq:Sol_Green_WPD_w1})-(\ref{eq:Sol_Green_WPD_w6}) which
present the solution of \eq(\ref{eq:G_equation}).  We now turn to the
discussion of these results.

\section{Discussion}
\label{sec:discussion}

\begin{figure}
\centering
\includegraphics[width=\linewidth]{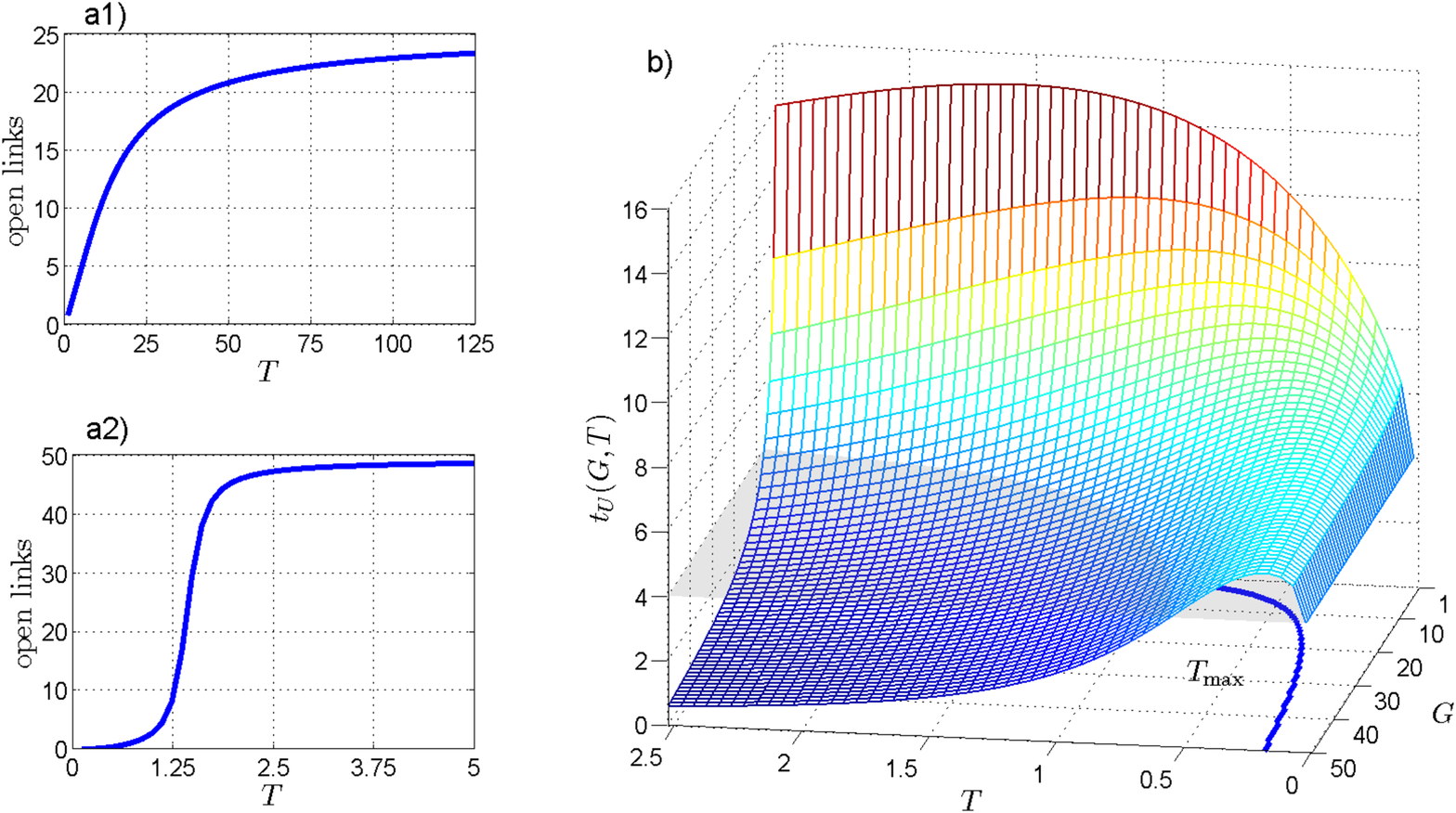}
\caption{Left two panels: Mean number of open links in equilibrium for
  Kittel's molecular zipper as a function of the reservoir temperature
  $T$ for $N=50$ links, and (a1)) $G=1$ and (a2)) $G=2$.  Right panel
  b): Unfolding time $t_{U}$, \eq(\ref{eq:unfolding_time-2}), as a
  function of $T$ and $G$ for $T_0 =7.5$, $v=0.25$, and $N=2$.  Above
  (below) the horizontal plane in the graph, the opened (closed) state
  is energetically favored. In the base plane, the temperature $T_{\rm
    max}$ of maximal unfolding time in dependence of the degeneracy
  factor $G$ is shown.}
\label{fig:tUnfold}
\end{figure}

In Kittel's work \cite{Kittel_1969} the {\em equilibrium\/} properties
of the zipper are studied. The mean number of open links in equilibrium
always increases with increasing temperature, but the form of this
increase is different for the degeneracy factor $G=1$ (cf.\
\fig\ref{fig:tUnfold}a1)) and for $G>1$ (cf.\
\fig\ref{fig:tUnfold}a2)).  For $G=1$, the mean number of open links
increases smoothly with a concave curvature, while for $G>1$, the
curve exhibits a sharp increase in a narrow temperature interval and
resembles a first-order phase transition. In the following discussion
of representative results for the nonequilibrium dynamics and
energetics, we use $d$, $d/k_{\rm B}$, and $\nu^{-1}$ as units for
energy, temperature and time, respectively.

In the unidirectional unzipping process, the time evolution towards
the completely unzipped state is again sensitive to the degeneracy
factor $G$.  Let us define an unfolding time $t_U$ by the condition
that at $t=t_U$ the zipper has completely unfolded with a probability
of 99,9\%, i.e.\
\begin{equation}
p_N(t_U)= R_{N 1}(t_U) = 1 - \epsilon\,\,,
\label{eq:unfolding_time}
\end{equation}
where $\epsilon = 0.001$. With respect to the $N$ dependence of $t_U$
(and other quantities to be discussed below), we found that its
behavior is similar for $N=2$ and $N>2$, and we therefore restrict the
following discussion to the two-state case
$N=2$. \eq(\ref{eq:unfolding_time}) can then be inverted after
inserting $R_{N 1}(t_U)$ from \eq(\ref{eq:R_big}) in
\eq(\ref{eq:unfolding_time}), yielding
\begin{equation}
t_U = [1/(\beta v)]\,\ln\left[1 - (\beta v/\alpha)\ln\epsilon\right]\,\,.
\label{eq:unfolding_time-2}
\end{equation}

\begin{figure}
\includegraphics[width=\linewidth]{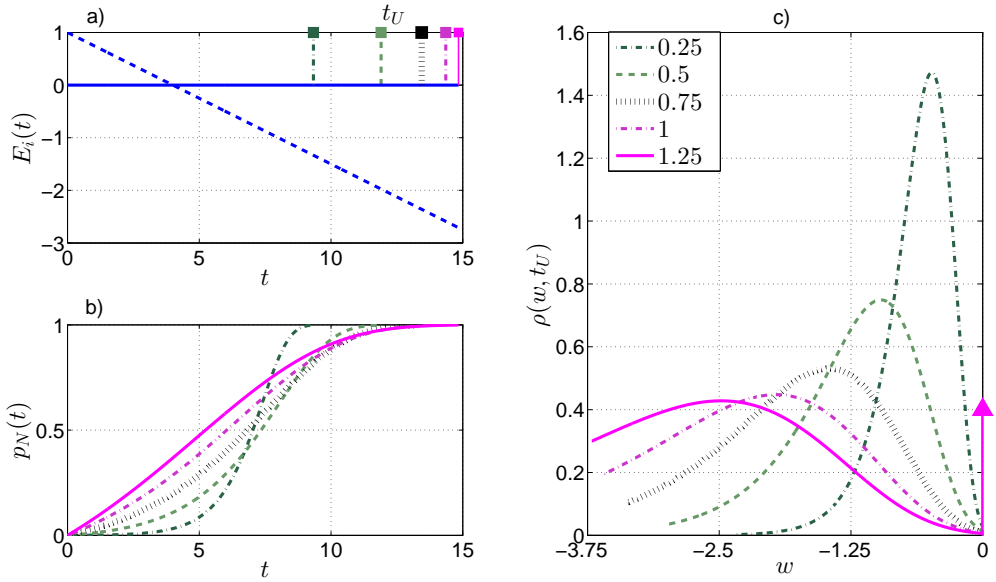}
\caption{Dynamics of a) the energy levels [\eq(\ref{eq:Energies})], b)
  the occupation probability $p_{N}(t)$ of the opened state
  [\eq(\ref{eq:Probabilities})], and c) the probability density
  $\rho(w,t_U)$ of the work [\eq(\ref{eq:WPD})] for $G=1$ and several
  values of the reservoir temperature $T$.  The other parameters are
  the same as in \fig\ref{fig:tUnfold}.  The assignment of the line
  styles to the temperature given in the legend of c) applies also to
  parts a) and b).  In a) we compare the dynamics of the two energy
  levels with the unfolding time $t_{U}$
  [\eq(\ref{eq:unfolding_time-2})] at different temperatures marked by
  the vertical lines. The full (dashed) line gives the energy $E_1(t)$
  ($E_2(t)$) of the closed (unzipped) state. The arrows in c)
  represent the weights and the positions of the $\delta$-functions,
  which form the singular part of the probability density.}
\label{fig:T_slow}
\end{figure}

\begin{figure}
\includegraphics[width=\linewidth]{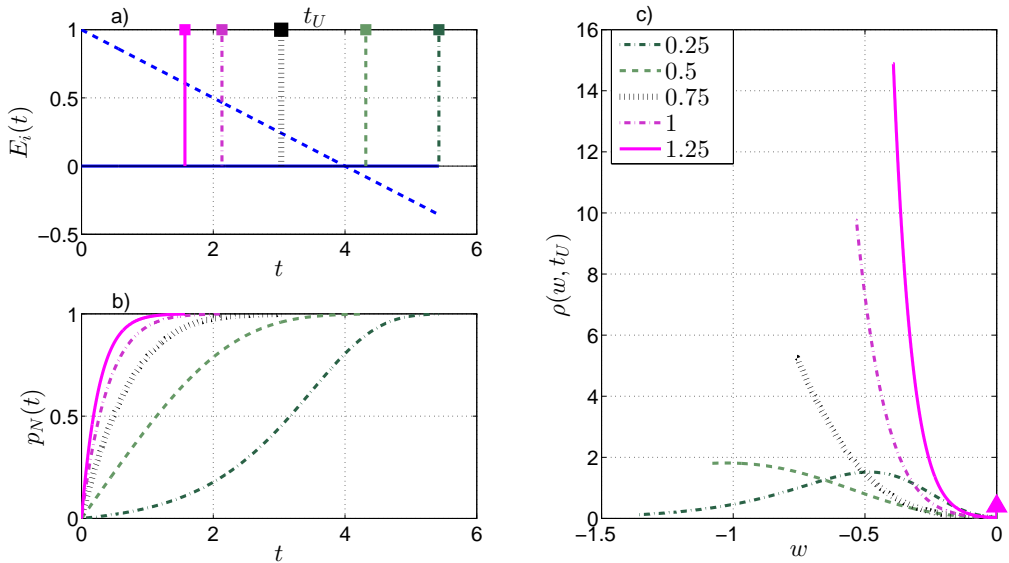}
\caption{Same quantities as in \fig\ref{fig:T_slow} for $G=50$ and
  otherwise the same set of parameters.}
\label{fig:T_fast}
\end{figure}

\begin{figure}
\includegraphics[width=\linewidth]{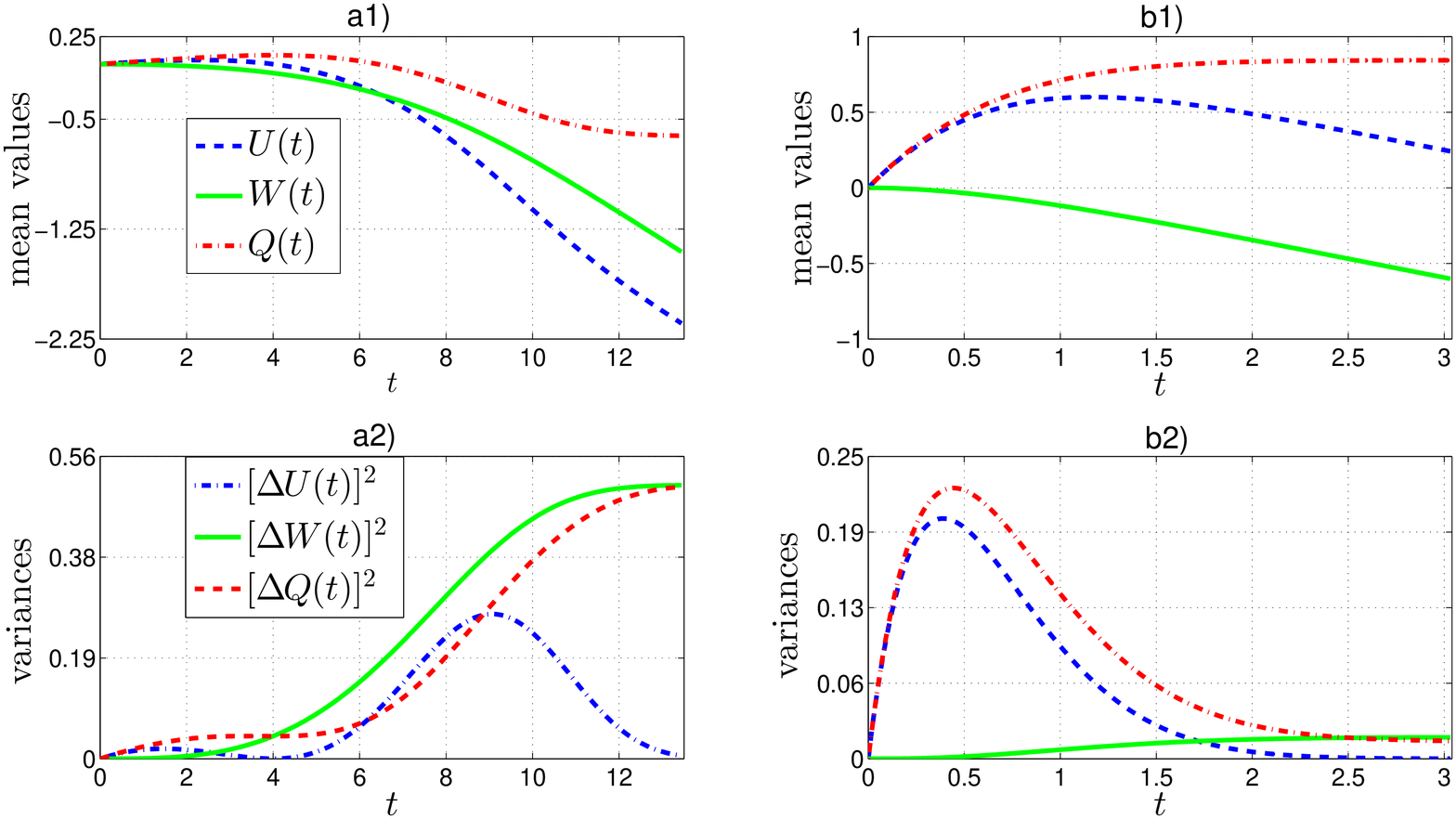}
\caption{Upper two panels: Mean value of the internal energy
  [\eq(\ref{eq:Umean})], work [\eq(\ref{eq:Wmean})] and heat
  [\eq(\ref{eq:Qmean})] as a function of time for a1) $G=1$ and b1)
  $G=50$. Lower two panels: Variances of internal energy
  [\eq(\ref{eq:Uvar})], work [\eq(\ref{eq:Wvar})] and heat
  [\eq(\ref{eq:Qvar})] as a function of time for a2) $G = 1$ and b2)
  $G = 50$. The temperature is $T = 0.75$ and the other parameters are
  the same as in \fig\ref{fig:tUnfold}.}
\label{fig:thermodynamics}
\end{figure}

For small temperatures [large argument of the exponential and/or small
prefactor $g$ in \eq(\ref{eq:rates})] the transitions are driven
predominantly by the magnitude of the energy gap between the closed
and the opened state. For large temperatures [small argument of the
exponential and/or large prefactor $g$ in \eq(\ref{eq:rates})] by
contrast, the dynamics is governed by the entropy difference between
the closed and the opened state, and hence by the degeneracy factor
$G$. The dependence of the unfolding time $t_U$ on $G$ and $T$ is
plotted in \fig\ref{fig:tUnfold}b) for a representative set of
parameters. For any fixed nonzero temperature, the unfolding time
decreases with increasing $G$, while for fixed $G$, the
temperature-dependence of the unfolding time exhibits a maximum at a
temperature $T=T_{{\rm max}}(G)$, where $T_{{\rm max}}(G)$ decreases
with increasing $G$, see \fig\ref{fig:tUnfold}b).

Let us now consider a certain temperature $T_0$ and call, for this
temperature, the {\em fast-unzipping regime\/} and {\em slow-unzipping
  regime\/} the ranges of $G$-values, where $T_{{\rm max}}(G)<T_0$ and
$T_{{\rm max}}(G)>T_0$, respectively.  In these two regimes the
dynamics and energetics of the molecular zipper exhibit a
qualitatively different behavior.  In particular we find (i) a
different time-dependence of the $N$-th state's occupation
probability, cf.\ \figs \ref{fig:T_slow}b) and \ref{fig:T_fast}b),
(ii) a different form of the curves describing the work needed to open
the zipper, cf.\ \figs \ref{fig:T_slow}c) and \ref{fig:T_fast}c),
(iii) a different mean value of heat accepted by the zipper during the
unzipping, cf.\ \figs \ref{fig:thermodynamics}a1) and
\ref{fig:thermodynamics}b1), and iv) different values of the variances
of the internal energy and heat during the unzipping, cf.\ \figs
\ref{fig:thermodynamics}a2) and \ref{fig:thermodynamics}b2). These
features will be now discussed in more detail.

\fig\ref{fig:T_slow} illustrates the slow-unzipping regime. The
probability $p_N(t)$ that the zipper has reached the opened state
until time $t$ first increases slowly. After the time $d/v$ the energy
of the opened state becomes lower than that of the closed one. This
leads to more frequent transitions and accordingly $p_N(t)$ increases
more rapidly. Notice that the curves exhibit a change of their second
derivative.  The work probability density (WPD) during the unzipping
has a maximum located inside its finite support, cf.\
\fig\ref{fig:T_slow}c). The value of the work at the left (right)
border of the support equals the work done on the zipper when it
dwells during the time interval $[0,t_{U}]$ in the opened (closed)
state.  From the position of the WPD peak we can conclude that, for a
typical trajectory of the stochastic process $\mathsf{D}(t)$, the work
consists of two comparable fractions. The first (second) part of the
work is performed while the system dwells in the closed (opened)
state.  At time $t_U$, $0.1\%$ of the trajectories will give molecules
still residing in the zipped state. These trajectories contribute to
the singular ($\delta$-function) components of WPDs, which are
depicted in \fig\ref{fig:T_slow}c) by the vertical arrows
\cite{Subrt_2007, Chvosta_2007, Einax_2009, Chvosta_2010}. Note that
the curves plotted for the temperatures $T=1$ and $T=1.25$, which are
close to the temperature $T_{{\rm max}}$(G) for $G=1$, cf.\
\fig\ref{fig:tUnfold}b), become similar to the curves for these
temperatures obtained in the fast-unzipping regime for $G=50$, see
\fig\ref{fig:T_fast}.

\fig\ref{fig:T_fast} illustrates the fast-unzipping regime. The
probability $p_N(t)$ rapidly increases from the very beginning of the
process, cf.\ \fig\ref{fig:T_fast}b), i.e.\ the zipper opens before
the opened state becomes energetically preferred. The maximum of the
WPD is located at the left border of its support, cf.\
\fig\ref{fig:T_fast}c). This means that, for the majority of the
trajectories, the substantial part of the work is done while the
system dwells in the opened state. Note that the curves plotted for
the temperature $T=0.25$, which is close to the boundary temperature
$T_{{\rm max}}$(G) for $G=50$, cf.\ \fig\ref{fig:tUnfold}b), become
similar to the curves for this temperature in the fast-unzipping
regime for $G=1$, see \fig\ref{fig:T_slow}.

\fig\ref{fig:thermodynamics} illustrates the dynamics of the
thermodynamic quantities (\ref{eq:Umean})--(\ref{eq:Qvar}) in the two
unzipping regimes.  For an arbitrary trajectory of $\mathsf{D}(t)$
which resides during the time interval $[t',t]$ in the $i$-th state,
the work performed on the system is $E_i(t) - E_i(t')$, cf.\
\eq(\ref{eq:work}).  In our model, the energies of the states decrease
linearly with time and accordingly the mean work is a monotonically
decreasing function of time, see \figs\ref{fig:thermodynamics}a1) and
\ref{fig:thermodynamics}b1). Heat is exchanged with the reservoir when
the molecule changes its state. It is absorbed by the molecule if the
transition brings the molecular zipper to a state with higher energy.
Since in our setting the transitions are unidirectional, the molecule
necessarily absorbs heat up to the time $t_{E}=d/v$, where the
energies of the states become the same, cf.\ \figs\ref{fig:T_slow}a)
and \ref{fig:T_fast}a). For times $t>t_{E}$, the molecule delivers
heat to the environment.  In the slow-unzipping regime we have $t_U >
t_{E}$. This implies that the mean heat first increases and then
decreases, cf.\ \fig\ref{fig:thermodynamics}a1). By contrast, in the
fast-unzipping regime where $t_U < t_{E}$, the mean heat monotonically
increases, cf.\ \fig\ref{fig:thermodynamics}b1).  Finally, due to the
transitions to the state with higher energy at the very beginning of
the process, cf.\ \figs\ref{fig:T_slow} and \ref{fig:T_fast}, the mean
internal energy (\ref{eq:Umean}) develops a single maximum.

The variances of $\mathsf{U}(t)$, $\mathsf{W}(t)$ and $\mathsf{Q}(t)$,
cf.\ \eqs(\ref{eq:Uvar})--(\ref{eq:Qvar}), are plotted in
\figs\ref{fig:thermodynamics}a2) and \figs\ref{fig:thermodynamics}b2).
If the variance of the internal energy approaches zero, the variance
of the work becomes equal to that of the heat.  All variances approach
constant values at large times. In fact, for $t>t_{U}$, almost all
trajectories have brought the molecule into the opened state.  As a
consequence, the increments to the work, heat and internal energy are
nearly constant during the time interval $[t_U, t]$.  Therefore the
form of their probability densities does not change, the curves just
move along the energy axis.

The probability density for the internal energy has no continuous
component. In general it consists of $N$ $\delta$-functions located at
the energies of the individual states. The corresponding weights are
given by the occupation probabilities of the states. In the
slow-unzipping regime, $[\Delta U(t)]^2$ vanishes at time $t=0$ and at
time $t_E=d/v$, when the state energies are equal, and it becomes very
close to zero for $t\gtrsim t_U$.  In between these time instants, the
variance develops a maximum, cf.\ \fig\ref{fig:thermodynamics}a2).  In
the fast-unzipping regime, the function $[\Delta U(t)]^2$ displays
just one maximum, cf.\ \fig\ref{fig:thermodynamics}b2), because $t_U <
t_{E}$.

In both the fast and slow unzipping regime, the variance of the work
monotonically increases and approaches a constant for large times,
cf.\ \fig\ref{fig:thermodynamics}a2) and
\fig\ref{fig:thermodynamics}b2).  The absolute value of the first
derivate of the variance $[\Delta Q(t)]^2$ is given by the product of
the two quantities, the energy difference between the closed and the
opened states at time $t$ and the probability that the zipper opens at
the time $t$, which were shown in \figs\ref{fig:T_slow} and
\ref{fig:T_fast}.  In the slow-unzipping regime, the majority of the
transitions occurs during the time interval $[t_E,t_U]$. During this
time interval, the energy difference between the states monotonically
increases. In the fast-unzipping regime, by contrast, nearly all
transitions take place before time $t_E/2$. The sooner the transition,
the larger is the amount of transferred heat. As a result, at small
times, the heat probability density has one peak at a large value,
coming from the trajectories with a transition to the opened state,
and another peak at zero heat exchange, originating from trajectories
without a transition.  With increasing time the amount of the
trajectories with a transition to the opened state increases rapidly,
and the peak close to zero heat moves towards the peak at a large heat
value. This explains the behavior of the heat variances in
\figs\ref{fig:thermodynamics}a2) and \ref{fig:thermodynamics}b2).

Finally, one may ask how our results are affected if the variation
(``static disorder") in base pairing energies is included in the
modeling. To this end we have performed Monte Carlo simulations
\cite{Holubec_2011} of the stochastic process for a molecular zipper
with $N=10$ states, where the initial energies $E_k(0)$,
$k=1,\ldots,N$, in \eq(\ref{eq:Energies}) are given by
$E_k(0)=(k-1)(\Delta+\eta_k)$, corresponding to different losses of
energies due to variations in base pair bondings.  The $\eta_k$ were
chosen as random numbers from a box distribution in the interval
$[-\Delta/3,\Delta/3]$. Results from these simulations for a number of
realizations of this disorder were compared to the predictions of the
analytical theory for the "ordered case", where
$E_k(0)=(k-1)\Delta$. We found that, for typical realizations of sets
of $\eta_k$, the shapes of the work distributions are very similar,
while the peak positions and peak heights are shifted slightly. Also
the probability distributions $p_N(t)$ for the zipper to fully open
until time $t$ are, for these sets of $\eta_k$, very similar in
shape. Analogous to the peak positions of the work distributions, the
onset of opening shifts slightly from realization to realization. The
shifts of the peaks and of the onset of the opening are controlled by
the largest base pair bonding (largest $\eta_k$), which governs the
unfolding time.

\ack Support of this work by the Ministry of Education of the Czech
Republic (project No.\ MSM 0021620835), by the Grant Agency of the
Charles University (grant No.\ 143610, and grant No.\ 301311), by the
Charles University in Prague (project No.\ SVV-2012-265 301), and by
the Deutsche Akademische Austauschdienst (DAAD, project No.\
MEB101104) is gratefully acknowledged.


\begin{thebibliography}{99}
\bibitem{comm:gamma-dfb} Smaller (larger) $\gamma$ would correspond to
  saddle points lying closer to state $k$ ($k+1$) in configuration
  space. The derivations in Secs.~\ref{sec:evolution} and
    \ref{sec:solution} can be performed analogously for any $\gamma$
    and $F_b$.
\bibitem{comm:modif-jarz} In this case one can show that
  $\langle\exp(-\beta w)\rangle=\phi(t)\exp\{-\beta[F(t)-F(1,0)]\}
  =\phi(t)\exp[-\beta F(t)]$, where $F(1,0)=0$ from
  \eq(\ref{eq:Energies}) and $F(t)= -\beta^{-1}\ln[(A^N-1)/(A-1)]$,
  $A=\exp[-\beta(\Delta+vt)]/G$, is the free energy for an equilibrated
  system with state energies $F(k,t)$ (protocol variables) at time
  $t$; $\phi(t)$ is the probability for the zipper to be in the
  fully closed state $k=1$ under the reversed protocol, if the states
  are initially distributed according to the equilibrium distribution
  $\exp[F(t)-F(k,t)]$.
\end{thebibliography}
\end{document}